# Optimal scaling of the random walk Metropolis on elliptically symmetric unimodal targets

CHRIS SHERLOCK[1] and GARETH ROBERTS[2]

[1]*Department of Mathematics and Statistics, Lancaster University, Lancaster, LA1 4YF, UK.*
*E-mail: c.sherlock@lancaster.ac.uk*
[2]*Department of Statistics, University of Warwick, Coventry, CV4 7AL, UK.*
*E-mail: gareth.o.roberts@warwick.ac.uk*

Scaling of proposals for Metropolis algorithms is an important practical problem in MCMC implementation. Criteria for scaling based on empirical acceptance rates of algorithms have been found to work consistently well across a broad range of problems. Essentially, proposal jump sizes are increased when acceptance rates are high and decreased when rates are low. In recent years, considerable theoretical support has been given for rules of this type which work on the basis that acceptance rates around 0.234 should be preferred. This has been based on asymptotic results that approximate high dimensional algorithm trajectories by diffusions. In this paper, we develop a novel approach to understanding 0.234 which avoids the need for diffusion limits. We derive explicit formulae for algorithm efficiency and acceptance rates as functions of the scaling parameter. We apply these to the family of elliptically symmetric target densities, where further illuminating explicit results are possible. Under suitable conditions, we verify the 0.234 rule for a new class of target densities. Moreover, we can characterise cases where 0.234 fails to hold, either because the target density is too diffuse in a sense we make precise, or because the eccentricity of the target density is too severe, again in a sense we make precise. We provide numerical verifications of our results.

## 1. Introduction

The Metropolis–Hastings updating scheme provides a very general class of algorithms for obtaining a dependent sample from a target distribution, $\pi(\cdot)$. Given the current value $\mathbf{X}$, a new value $\mathbf{X}^*$ is proposed from a pre-specified Lebesgue density $q(\mathbf{x}^*|\mathbf{x})$ and is then accepted with probability $\alpha(\mathbf{x}, \mathbf{x}^*) = \min\left(1, (\pi(\mathbf{x}^*)q(\mathbf{x}|\mathbf{x}^*))/(\pi(\mathbf{x})q(\mathbf{x}^*|\mathbf{x}))\right)$. If the proposed value is accepted it becomes the next current value ($\mathbf{X}' \leftarrow \mathbf{X}^*$), otherwise the current value is left unchanged ($\mathbf{X}' \leftarrow \mathbf{X}$).

Consider the $d$-dimensional **random walk Metropolis** (RWM) [7]:

$$q(\mathbf{x}^*|\mathbf{x}) = \frac{1}{\lambda^d} r\left(\frac{\mathbf{x}^* - \mathbf{x}}{\lambda}\right) = \frac{1}{\lambda^d} r\left(\frac{\mathbf{y}^*}{\lambda}\right), \qquad (1)$$

where $\mathbf{y}^* := \mathbf{x}^* - \mathbf{x}$ is the proposed jump, and $r(\mathbf{y}) = r(-\mathbf{y})$ for all $\mathbf{y}$. In this case the acceptance probability simplifies to

$$\alpha(\mathbf{x}, \mathbf{x}^*) = \min\left(1, \frac{\pi(\mathbf{x}^*)}{\pi(\mathbf{x})}\right). \qquad (2)$$

Now consider the behaviour of the RWM as a function of the scale of proposed jumps, $\lambda$, and some measure of the scale of variability of the target distribution, $\eta$. If $\lambda \ll \eta$ then, although proposed jumps are often accepted, the chain moves slowly and exploration of the target distribution is relatively inefficient. If $\lambda \gg \eta$ then many proposed jumps are not accepted, the chain rarely moves and exploration is again inefficient. This suggests that given a particular target and form for the jump proposal distribution, there may exist a finite scale parameter for the proposal with which the algorithm will explore the target as efficiently as possible. We are concerned with the definition and existence of an optimal scaling, its asymptotic properties and the process of finding it. We start with a brief review of current literature on the topic.

### 1.1. Existing results for optimal scaling of the RWM

Existing literature on this problem has concentrated on obtaining a limiting diffusion process from a sequence of Metropolis algorithms with increasing dimension. The speed of this limiting diffusion is then maximised with respect to a transformation of the scale parameter to find the optimally scaled algorithm. Roberts *et al.* [9] first follow this program for densities of the form

$$\pi(\mathbf{x}) = \prod_{i=1}^{d} f(x_i) \qquad (3)$$

using Gaussian jump proposals, $\mathbf{Y}^{(d)} \sim N(\mathbf{0}, \sigma_d^2 \mathbf{I}_d)$. Here and throughout this article $\mathbf{I}_d$ denotes the $d$-dimensional identity matrix. For high dimensional targets which satisfy certain moment conditions it is shown that the optimal value of the scale parameter satisfies $d^{1/2}\hat{\lambda}_d = l$, for some fixed $l$ which is dependent on the roughness of the target. Particularly appealing, however, from a practical perspective, is the following distribution-free interpretation of the optimal scaling for the class of distributions given by (3). It is the scaling that leads to the proportion 0.234 of proposed moves being accepted.

Empirically this "0.234" rule has been observed to be approximately right much more generally. Extensions and generalisations of this result can be found in Roberts and Rosenthal [10], which also provides an accessible review of the area, and Bedard [2], Breyer and Roberts [3], Roberts [8]. The focus of much of this work is in trying to characterise when the "0.234" rule holds and to explain how and why it breaks down in other situations.

One major disadvantage of the diffusion limit work is its reliance on asymptotics in the dimensionality of the problem. Although it is often empirically observed that the limiting behaviour can be seen in rather small dimensional problems (see, e.g., Gelman *et al.* [5]) it is difficult to quantify this in any general way.

In this paper we adopt a finite dimensional approach, deriving and working with explicit solutions for algorithm efficiency and overall acceptance rates.

### 1.2. Efficiency and expected acceptance rate

In order to consider the problem of optimising the algorithm, an optimisation criterion needs to be chosen. Unfortunately this is far from unique. In practical MCMC, interest may lie in the estimation of a collection of expected functionals. For any one of these functionals, $f$ say, a plausible criterion to minimise is the stationary integrated autocorrelation time for $f$ given by

$$\tau_f = 1 + 2\sum_{i=1}^{\infty} \mathrm{Cor}(f(X_0), f(X_i)).$$

Under appropriate conditions, the MCMC central limit theorem for $\{f(X_i)\}$ gives a Monte Carlo variance proportional to $\tau_f$. This approach has two major disadvantages. First, estimation of $\tau_f$ is notoriously difficult, and second, this optimisation criterion gives a different solution for the "optimal" chain for different functionals $f$.

In the diffusion limit, the problem of non-uniqueness of the optimal chain is avoided since in all cases $\tau_f$ is proportional to the inverse of the diffusion speed. This suggests that plausible criteria might be based on optimising properties of single increments of the chain.

The most general target distributions that we shall examine here possess elliptical symmetry. If a $d$-dimensional target distribution has elliptical contours then there is a simple invertible linear transformation $\mathbf{T} : \Re^d \to \Re^d$ which produces a spherically symmetric target. To fix it (up to an arbitrary rotation) we define $\mathbf{T}$ to be the transformation that produces a spherically symmetric target with unit scale parameter. Here the exact meaning of "unit scale parameter" may be decided arbitrarily or by convention. The scale parameter $\beta_i$ along the $i$th principal axis of the ellipse is the $i$th eigenvalue of $\mathbf{T}^{-1}$.

Let $\mathbf{X}$ and $\mathbf{X}'$ be consecutive elements of a stationary chain exploring a $d$-dimensional target distribution. A natural efficiency measure for elliptical targets is Mahalanobis distance, for example, Krzanowski [6]:

$$S_d^2 := \mathbb{E}[\|\mathbf{X}' - \mathbf{X}\|_\beta^2] := \mathbb{E}\left[\sum_{i=1}^d \frac{1}{\beta_i^2}(X_i' - X_i)^2\right] = \mathbb{E}\left[\sum_{i=1}^d \frac{1}{\beta_i^2} Y_i^2\right], \tag{4}$$

where $X_i'$ and $X_i$ are the components of $\mathbf{X}'$ and $\mathbf{X}$ along the $i$th principal axis and $Y_i$ are components of the realised jump $\mathbf{Y} = \mathbf{X}' - \mathbf{X}$. We refer to this as the expected square jump distance, or ESJD. We will relate ESJD to expected acceptance rate (EAR) which we define as $\overline{\alpha}_d := E[\alpha(\mathbf{X}, \mathbf{X}^*)]$, where the expectation is with respect to the joint law for the current value $\mathbf{X}$ and the proposed value $\mathbf{X}^*$. Note that we are not interested in the value of the ESJD itself but only in the scaling and EAR at which the maximum ESJD is attained.

### 1.3. Outline of this paper

The body of this paper investigates the RWM algorithm on spherically and then elliptically symmetric unimodal targets. Section 2 considers finite dimensional algorithms on spherically symmetric unimodal targets and derives explicit formulae for ESJD and EAR in terms of the scale parameter associated with the proposed jumps (Theorem 1). Several example algorithms are then introduced and the forms of $\overline{\alpha}_d(\lambda)$ and $S_d^2(\lambda)$ are derived for specific values of $d$ either analytically or by numerical integration. Numerical results for the relationship between the optimal acceptance rate and dimension are then described; in most of these examples the limiting optimal acceptance rate appears to be *less than* 0.234.

The explicit formulae in Theorem 1 involve the target's marginal one-dimensional distribution function. Theorem 2 of Section 3 provides a limiting form for the marginal one-dimensional distribution function of a spherically symmetric random variable as $d \to \infty$ and Theorem 3 combines this with a result from measure theory to provide limiting forms for EAR and ESJD as $d \to \infty$. A natural next step would be to use the limiting ESJD to estimate a limiting optimal scale parameter rather than directly examining the limit of the optimal scale parameters of the finite dimensional ESJDs. It is shown that this process is sometimes invalid when the target contains a mixture of scales that produce local maxima in ESJD and whose ratio increases without bound. Exact criteria are provided in Lemma 2 and are related to the numerical examples.

Many "standard" sequences of distributions satisfy the condition that as $d \to \infty$ the probability mass becomes concentrated in a spherical shell which itself becomes infinitesimally thin relative to its radius. Thus the random walk on a rescaling of the target is, in the limit, effectively confined to the surface of this shell. Theorem 4 considers RWM algorithms on sequences of spherically symmetric unimodal targets where the sequence of proposal distributions satisfies this "shell condition". It is shown that if the target sequence also satisfies the "shell condition" then the limiting optimal EAR is 0.234; however, if the target mass does not converge to an infinitesimally thin shell then the limiting optimal EAR (if it exists) is strictly less than 0.234. Rescalings of both the target and proposal are usually required in order to stabilise the radius of the shell, whether or not it becomes infinitesimally thin. These influence the form of the optimal scale parameter so that in general it is not proportional to $d^{-1/2}$. Corollary 4 provides an explicit formula that is consistent with the numerical examples.

Section 4 extends the results for finite dimensional random walks to all elliptically symmetric targets. Limit results are extended through Theorem 5 to sequences of elliptically symmetric targets for which the ellipses do not become too eccentric. The article concludes in Section 5 with a discussion.

## 2. Exact results for finite dimension

In this section we derive Theorem 1, which provides exact formulae for ESJD and EAR for a random walk Metropolis algorithm acting on a unimodal spherically symmetric target. The formulae in Theorem 1 refer to the target's marginal one-dimensional distribution function; these are then converted to use the more intuitive marginal radial distribution function. Several example targets are introduced and results from exact calculations of ESJD and EAR are presented.

We adopt the notation outlined in Section 1.2. All distributions (target and proposal) are assumed to have densities with respect to Lebesgue measure, and we consider the chain to be stationary so that the marginal densities of both $\mathbf{X}$ and $\mathbf{X}'$ are $\pi(\cdot)$. We also assume that the space of possible values for element $\mathbf{x}$ of a $d$-dimensional chain is $\Re^d$.

We consider *only target densities with a single mode*; however, the density need not decrease with strict monotonicity and may have a series of plateaux. We refer to random variables with such densities as *unimodal*. In this section and the section that follows we further restrict our choice of target to include only random variables where the density has spherical contour lines. Such random variables are termed *isotropic* or *spherically symmetric*. ESJD is as defined in (4) where the expectation is taken with respect to the joint law for the current position and the realised jump. For a spherical target $\beta_i = \beta \forall i$, the ESJD is proportional to the expected squared Euclidean distance, and both are maximised by the same scaling $\hat{\lambda}$. Since the constant of proportionality, $\beta$, derives from an arbitrary definition of "unit scale parameter", we simply set it to 1 for spherically symmetric random variables.

Denote the one-dimensional marginal distribution function of a general $d$-dimensional target $\mathbf{X}^{(d)}$ along unit vector $\hat{\mathbf{y}}$ as $F_{1|d}(x)$. When $\mathbf{X}^{(d)}$ is spherically symmetric, this is independent of $\hat{\mathbf{y}}$, and we simply refer to it as *the* one-dimensional marginal distribution function of $\mathbf{X}^{(d)}$. The following is proved in Appendix A.1.

**Theorem 1.** *Consider a stationary random walk Metropolis algorithm on a spherically symmetric unimodal target which has marginal one-dimensional distribution function $F_{1|d}(x)$. Let jumps be proposed from a symmetric density as defined in (1). In this case the expected acceptance rate and the expected square jump distance are*

$$\overline{\alpha}_d(\lambda) = 2\mathbb{E}[F_{1|d}(-\tfrac{1}{2}\lambda|\mathbf{Y}|)] \quad and \tag{5}$$

$$S_d^2(\lambda) = 2\lambda^2 \mathbb{E}[|\mathbf{Y}|^2 F_{1|d}(-\tfrac{1}{2}\lambda|\mathbf{Y}|)], \tag{6}$$

*where the expectation is taken with respect to measure $r(\cdot)$.*

The marginal distribution function $F_{1|d}(-\lambda|\mathbf{Y}|/2)$ is bounded and decreasing in $\lambda$. Also $\lim_{x \to \infty} F_{1|d}(-x) = 0$ and by symmetry, provided $F_{1|d}(\cdot)$ is continuous at the origin, $\lim_{x \to 0} F_{1|d}(-x) = 0.5$. Applying the bounded convergence theorem to (5) we therefore obtain the following intuitive result:

**Corollary 1.** *Let $\lambda$ be the scaling parameter for any RWM algorithm on a unimodal isotropic target Lebesgue density. In this situation the EAR at stationarity $\overline{\alpha}_d(\lambda)$ decreases with increasing $\lambda$, with $\lim_{\lambda \to 0} \overline{\alpha}_d(\lambda) = 1$ and $\lim_{\lambda \to \infty} \overline{\alpha}_d(\lambda) = 0$.*

In our search for an optimal scaling there is an implicit assumption that such a scaling exists. This was justified intuitively in Section 1 but the existence of an optimal scaling has previously only been proven for the limiting diffusion process as $d \to \infty$; see Roberts *et al.* [9]. Starting from Theorem 1 the following is relatively straightforward to prove (see Sherlock [11]) and starts to justify a search for an optimal scaling for a finite dimensional random walk algorithm rather than a limit process.

**Corollary 2.** *Consider a spherically symmetric unimodal d-dimensional target Lebesgue density $\pi(\mathbf{x})$. Let $\pi(\cdot)$ be explored via an RWM algorithm with proposal Lebesgue density $\frac{1}{\lambda^d} r(\mathbf{y}/\lambda)$. If $\mathbb{E}_\pi[|\mathbf{X}|^2] < \infty$ and $\mathbb{E}_r[|\mathbf{Y}|^2] < \infty$ then the ESJD of the Markov chain at stationarity attains its maximum at a finite non-zero value (or values) of $\lambda$.*

For the remainder of this section we examine the behaviour of real, finite dimensional examples of random walk algorithms. As well as being of interest in its own right, this will motivate Section 3 where Theorem 1 will provide the basis from which properties of EAR and ESJD are obtained as dimension $d \to \infty$. To render Theorem 1 of more use for practical calculation, we first convert it to involve the more intuitive marginal radial distribution rather than the marginal one-dimensional distribution function.

We introduce some further notation; write $\overline{F}_d(\cdot)$ and $\overline{f}_d(\cdot)$ for the marginal radial distribution and density functions of $d$-dimensional spherically symmetric target $\mathbf{X}^{(d)}$; these are the distribution and density functions of $|\mathbf{X}^{(d)}|$. The density of $|\mathbf{Y}|$ (when $\lambda = 1$) is denoted $\overline{r}_d(\cdot)$.

We start with a form for the one-dimensional marginal distribution function of a spherically symmetric random variable in terms of its marginal radial distribution function. Derivation of this result from first principles is straightforward; see Sherlock [11].

**Lemma 1.** *For any d-dimensional spherically symmetric random variable with continuous marginal radial distribution function $\overline{F}_d(r)$ with $\overline{F}_d(0) = 0$, the one-dimensional marginal distribution function along any axis is*

$$F_{1|d}(x_1) = \frac{1}{2}\left(1 + \text{sign}(x_1)\mathbb{E}_{X^{(d)}}\left[G_d\left(\frac{|x_1|^2}{|\mathbf{X}^{(d)}|^2}\right)\right]\right) \qquad (d \geq 1), \tag{7}$$

*where $\text{sign}(x) = 1$ for $x \geq 0$ and $\text{sign}(x) = -1$ for $x < 0$, and $G_d(\cdot)$ is the distribution function of $U_d$, with $U_1 = 1$ and*

$$U_d \sim Beta\left(\frac{1}{2}, \frac{d-1}{2}\right) \qquad (d > 1).$$

For the RWM we are concerned only with targets with Lebesgue densities. In this case both the marginal one-dimensional and radial distribution functions are continuous, and $\overline{F}_d(0) = 0$ as there can be no point mass at the origin (or anywhere else). Substituting (7) into (5) and (6) gives

$$\overline{\alpha}_d(\lambda) = \mathbb{E}_{\mathbf{Y}, \mathbf{X}^{(d)}}\left[K_d\left(\frac{\lambda|\mathbf{Y}|}{2|\mathbf{X}^{(d)}|}\right)\right] \quad \text{and} \quad S_d^2(\lambda) = \lambda^2 \mathbb{E}_{\mathbf{Y}, \mathbf{X}^{(d)}}\left[|\mathbf{Y}|^2 K_d\left(\frac{\lambda|\mathbf{Y}|}{2|\mathbf{X}^{(d)}|}\right)\right],$$

where $\mathbf{Y}$ is a random variable with density $r(\cdot)$ and $K_d(x) := 1 - G_d(x^2)$. The expectations depend on $\mathbf{X}$ and $\mathbf{Y}$ only through their moduli, thus allowing expressions for EAR and ESJD in terms of simple double integrals involving the marginal radial densities of $|\mathbf{X}|$ and $|\mathbf{Y}|$. For unimodal spherically symmetric targets we therefore obtain:

$$\overline{\alpha}_d(\lambda) = \int_0^\infty dy \int_{\lambda y/2}^\infty dx\, \overline{r}_d(y) \overline{f}_d(x) K_d\left(\frac{\lambda y}{2x}\right), \tag{8}$$

$$S_d^2(\lambda) = \lambda^2 \int_0^\infty dy \int_{\lambda y/2}^\infty dx\, \overline{r}_d(y) \overline{f}_d(x) y^2 K_d\left(\frac{\lambda y}{2x}\right). \tag{9}$$

Since $\mathbf{X}^{(d)}$ is spherically symmetric $\overline{f}_d(|\mathbf{x}|) = a_d |\mathbf{x}|^{d-1} \pi(\mathbf{x})$, where $a_d := 2\pi^{d/2}/\Gamma(d/2)$; see [1], Chapter 15. In the examples below we also consider only spherically symmetric proposals so that $\overline{r}_d(|\mathbf{y}|) = a_d |\mathbf{y}|^{d-1} r_d(\mathbf{y})$.

## 2.1. Explicit and computational results

Using (8) and (9) we first examine the dependency of EAR and ESJD on $\lambda$ for any given dimension. We then examine the behaviour of the optimal scaling and optimal acceptance rate as dimension $d$ increases.

Now $K_1(u) = 1$ for $u < 1$ and $K_1(u) = 0$ otherwise, and so for one-dimensional RWM algorithms the integrals in (8) and (9) may sometimes be evaluated exactly. For example, with a Gaussian target and Gaussian proposal, (8) and (9) give

$$\overline{\alpha}_1(\lambda) = \frac{2}{\pi} \tan^{-1}\left(\frac{2}{\lambda}\right) \quad \text{and}$$

$$S_1^2(\lambda) = \frac{2\lambda^2}{\pi}\left(\tan^{-1}\left(\frac{2}{\lambda}\right) - \frac{2\lambda}{\lambda^2 + 4}\right). \tag{10}$$

Maximising (10) numerically gives an optimal scaling of $\hat{\lambda} \approx 2.43$ which corresponds to an optimal EAR of 0.439.

With both target and proposal following a double exponential distribution, (8) and (9) produce

$$\overline{\alpha}_1(\lambda) = \frac{2}{\lambda + 2} \quad \text{and} \quad S_1^2(\lambda) = \frac{16\lambda^2}{(\lambda + 2)^3}.$$

$S_1^2$ and $\overline{\alpha}_1$ are thus related by the simple analytical expression $S_1^2 = 8\overline{\alpha}_1(1 - \overline{\alpha}_1)^2$, and the ESJD attains a maximum at an EAR of $1/3$, for which $\hat{\lambda} = 4$.

We now consider two example targets with $d = 10$: first, a simple Gaussian ($\pi_d(\mathbf{x}) \propto \mathrm{e}^{-|\mathbf{x}|^2/2}$), and second, a mixture of Gaussians:

$$\pi_d(\mathbf{x}) \propto (1-p_d)\mathrm{e}^{-|\mathbf{x}|^2/2} + p_d \frac{1}{d^d}\mathrm{e}^{-|\mathbf{x}|^2/(2d^2)} \quad (d \geq 2), \tag{11}$$

with $p_d = 1/d^2$. Both targets are explored using spherically symmetric Gaussian proposals; results are shown in Figure 1. As with the previous two examples, increasing $\lambda$ from 0 to $\infty$ decreases the EAR from 1 to 0, as deduced in Corollary 1. Further, in all four examples, as noted in Corollary 2, ESJD achieves a global maximum at finite, strictly positive values of $\lambda$. In the first three examples ESJD as a function of the scaling shows a single maximum; however, in the mixture example similar high ESJDs are achieved with two very different scale parameters (approximately 0.8 and 7.6). The acceptance rates at these maxima are 0.26 and 0.0026, respectively. The values $\hat{\lambda} = 0.8$ and $\hat{\alpha} = 0.26$ are almost identical to the optimal values for exploring a standard ten-dimensional Gaussian and so are ideal for exploring the first component of the mixture. Optimal exploration of the second component is clearly to be achieved by increasing the scale parameter by a factor of 10; however, the second component has a mixture weight of 0.01 and so the acceptance rate for such proposals is reduced accordingly. The mixture weighting of the second component, $1/d^2$, is just sufficient to balance the increase in optimal jump size for that component, with the result that the two peaks in ESJD are of equal heights.

We next examine the behaviour of the optimal scaling and the corresponding EAR as $d$ increases. Calculations are performed for eight different targets:

1. Gaussian density: $\pi_d(\mathbf{x}) \propto \mathrm{e}^{-|\mathbf{x}|^2/2}$;
2. exponential density: $\pi_d(\mathbf{x}) \propto \mathrm{e}^{-|\mathbf{x}|}$;
3. target with a Gaussian marginal radial density: $\pi_d(\mathbf{x}) \propto |\mathbf{x}|^{-d+1}\mathrm{e}^{-|\mathbf{x}|^2/2}$;
4. target with an exponential marginal radial density: $\pi_d(\mathbf{x}) \propto |\mathbf{x}|^{-d+1}\mathrm{e}^{-|\mathbf{x}|}$;
5. lognormal density altered so as to be unimodal:

$$\pi_d(\mathbf{x}) \propto \mathbb{1}_{\{|\mathbf{x}| \leq \mathrm{e}^{-(d-1)}\}} + \mathrm{e}^{-(\log|\mathbf{x}|+(d-1))^2/2}\mathbb{1}_{\{|\mathbf{x}| > \mathrm{e}^{-(d-1)}\}};$$

6. the mixture of Gaussians given by (11) with $p_d = 0.2$;
7. the mixture of Gaussians given by (11) with $p_d = 1/d$;
8. the mixture of Gaussians given by (11) with $p_d = 1/d^3$.

Proposals are generated from a Gaussian density. For each combination of target and proposal simple numerical routines are employed to find the scaling $\hat{\lambda}$ that produces the largest ESJD. Substitution into (8) gives the corresponding optimal EAR $\hat{\alpha}$.

Figure 2 shows plots of optimal EAR against dimension for example targets 1–4. The first of these is entirely consistent with Figure 4 in Roberts and Rosenthal [10], which shows optimal acceptance rates obtained through repeated runs of the RWM algorithm. The first two are consistent with a conjecture that the optimal EAR approaches 0.234 as $d \to \infty$; however, for examples targets 3 and 4, the optimal EAR appears to approach limits of approximately 0.10 and 0.06, respectively.

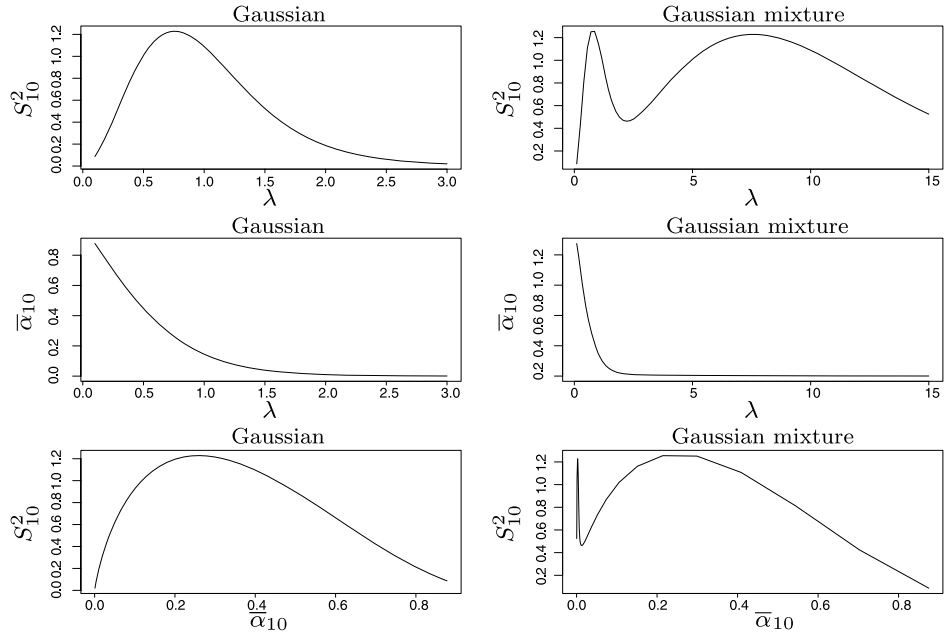

**Figure 1.** Plots for a Gaussian target (left) and the Gaussian mixture target of (11) with $p_d = 1/d^2$ (right), both at $d = 10$ and with a Gaussian jump proposal. Panels from top to bottom are (i) ESJD against scaling, (ii) EAR against scaling and (iii) ESJD against EAR.

For target 5 with $d = 1, 2$ or 3, plots of ESJD against scale parameter, EAR against scale parameter and ESJD against EAR (not shown) are heuristically similar to those for the standard Gaussian target in Figure 1. However, for $d = 1$, 2 and 3 the optimal EARs are approximately 0.111, 0.010 and 0.00057, respectively, and appear to be approaching a limiting optimal acceptance rate of 0.

Figure 3 shows plots of EAR against dimension for the three mixture targets (6–8). Here the asymptotically optimal EAR appears to be approximately 0.234/5, 0 and 0.234, respectively. The limiting behaviour of each of these examples is explained in the next section.

## 3. Limit results for spherically symmetric distributions

Theorem 1 provides exact analytical forms for the EAR and ESJD of an RWM algorithm on a unimodal spherically symmetric target in terms of the target's marginal one-dimensional distribution function. In this section we investigate the behaviour of EAR and ESJD in the limit as dimension $d \to \infty$. As groundwork for this investigation we must first examine the possible limiting forms of the marginal one-dimensional dis-tribution function of a spherically symmetric random variable. We adopt the following notation: Convergence in distribution is denoted by $\xrightarrow{D}$; convergence in probability is denoted by $\xrightarrow{p}$ and convergence in mean square by $\xrightarrow{m.s.}$.

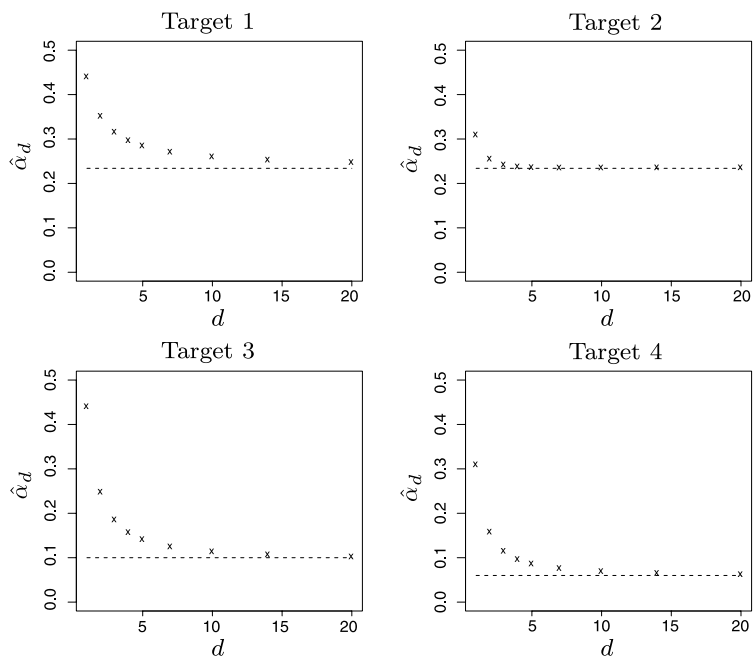

**Figure 2.** Plots of the optimal EAR $\hat{\alpha}$ against dimension for example targets 1–4 using a Gaussian jump proposal. The horizontal dotted line approximates the apparent asymptotically optimal acceptance rate of 0.234 in the first two plots and 0.10 and 0.06 in the third and fourth plots, respectively.

Convergence of the sequence of characteristic functions of a sequence of $d$-dimensional isotropic random variables (indexed by $d$) to that of a mixture of normals is proved as Theorem 2.21 of Fang *et al.* [4]. Thus the limiting marginal distribution along any given axis may be written as $X_1 = RZ$ with $Z$ a standard Gaussian and $R$ the mixing distribution. Sherlock [11] proves from first principles the following extension.

**Theorem 2.** *Let $\mathbf{X}^{(d)}$ be a sequence of $d$-dimensional spherically symmetric random variables. If there is a $k_d$ such that $|\mathbf{X}^{(d)}|/k_d \xrightarrow{D} R$ then the sequence of marginal one-dimensional distributions of $\mathbf{X}^{(d)}$ satisfies*

$$F_{1|d}\left(\frac{k_d}{d^{1/2}}x_1\right) \to \Theta(x_1) := \mathbb{E}_R\left[\Phi\left(\frac{x_1}{R}\right)\right],$$

*where $\Phi(\cdot)$ is the standard Gaussian distribution function.*

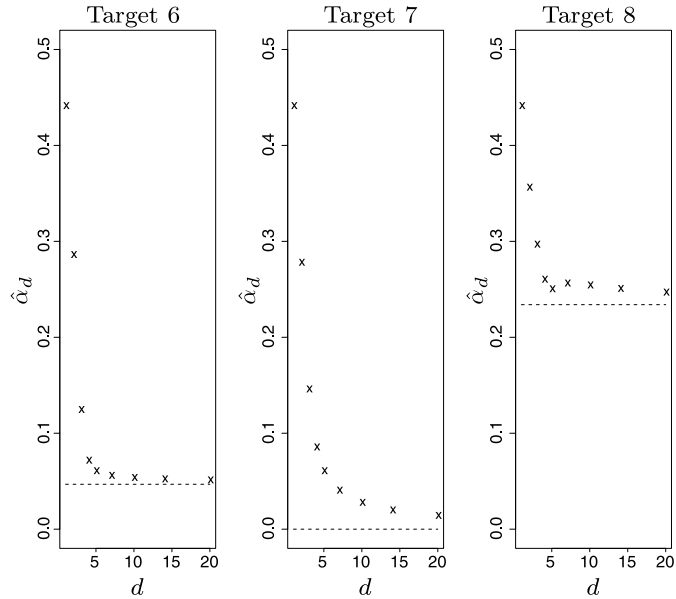

**Figure 3.** Plots of the optimal EAR $\hat{\alpha}$ against dimension for example targets 6–8 using a Gaussian jump proposal. The horizontal dotted line in each plot represents the apparent asymptotically optimal acceptance rate of 0.234/5, 0 and 0.234, respectively.

$|\mathbf{X}^{(d)}|$ possesses a Lebesgue density and therefore no point mass at the origin; however, the rescaled limit $R$ may possess such a point mass. Provided $R$ has no point mass at 0, the limiting marginal one-dimensional distribution function $\Theta(x_1)$ as defined in Theorem 2 is therefore continuous for all $x \in \Re$. This continuity implies that the limit in Theorem 2 is approached uniformly in $x_1$, and for this reason the lack of a radial point mass at 0 is an essential requirement in Theorem 3.

The condition of convergence of the rescaled modulus to 1 or to random variable $R$ will turn out to be the key factor in determining the behaviour of the optimal EAR as $d \to \infty$; we now examine this limiting convergence behaviour in more detail.

For many standard sequences of density functions there is a $k_d$ such that $|\mathbf{X}^{(d)}|/k_d \xrightarrow{p} 1$. This includes example targets 1 and 2 from Section 2.1, and more generally any density of the form $\pi_d(\mathbf{x}) \propto |\mathbf{x}|^a e^{-|\mathbf{x}|^c}$. An intuitive understanding of target sequences satisfying this condition is that, as $d \to \infty$ the probability mass becomes concentrated in a spherical shell which itself becomes infinitesimally thin relative to its radius. The random walk on a rescaling of the target is, in the limit, effectively confined to the surface of this shell.

Example targets 3 and 4 have marginal radial distributions which are always respectively a positive unit Gaussian and a unit exponential. The first term in the density of example target 5 simply ensures unimodality and becomes increasingly unimportant as $d$ increases. Trivial algebraic rearrangement of the second component shows that its marginal radial distribution has the same log-normal form whatever the dimension. Example targets 6–8 are examined in detail in Section 3.2.

### 3.1. A limit theorem for EAR and ESJD

We now return to the RWM and derive limiting forms for ESJD and EAR on unimodal spherically symmetric targets as $d \to \infty$. Henceforth it is assumed that the radial distribution of the target, rescaled by a suitable quantity $k_x^{(d)}$, converges weakly to some continuous limiting distribution, that of a random variable $R$. From Theorem 2, the limiting marginal distribution function $\Theta(\cdot)$ is in general a scaled mixture of Gaussian distribution functions but in the special case that $R$ is a point mass at 1 the scaled mixture of Gaussians clearly reduces to the standard Gaussian cumulative distribution function $\Phi(\cdot)$; $F_{1|d}(\frac{k_d}{d^{1/2}} x_1) \to \Phi(x_1)$.

Consider a sequence of jump proposal random variables $\{\mathbf{Y}^{(d)}\}$ with unit scale parameter. If there exist $k_y^{(d)}$ such that $|\mathbf{Y}^{(d)}|/k_y^{(d)}$ converges (in a sense to be defined) then simple limit results are possible. Implicit in the derivation of these limit results is a transformation of our target and proposal: $\tilde{\mathbf{X}}^{(d)} \leftarrow \mathbf{X}^{(d)}/k_x^{(d)}$ and $\tilde{\mathbf{Y}}^{(d)} \leftarrow \mathbf{Y}^{(d)}/k_y^{(d)}$. We define a **transformed scale parameter**

$$\mu_d := \frac{1}{2} \frac{d^{1/2} k_y^{(d)}}{k_x^{(d)}} \lambda_d. \tag{12}$$

A random walk on target density $(k_x^{(d)})^d \pi_d(k_x^{(d)} \mathbf{x})$ using proposal density $(k_y^{(d)})^d r_d(k_y^{(d)} \mathbf{y})$ and scale parameter $2\mu_d$ is therefore equivalent to a random walk on $\pi_d(\mathbf{x})$ using proposal $r_d(\mathbf{y})$ and a scale parameter $l = d^{1/2} \lambda_d$, a quantity which is familiar from the diffusion-based approach to optimal scaling (see Section 1.1). The following theorem characterises the limiting behaviour for EAR and ESJD for fixed values, $\mu$, of the transformed scale parameter; it is proved in Appendix A.2.

**Theorem 3.** *Let $\{\mathbf{X}^{(d)}\}$ be a sequence of $d$-dimensional unimodal spherically symmetric target random variables and let $\{\mathbf{Y}^{(d)}\}$ be the corresponding sequence of jump proposals. If there exist $\{k_x^{(d)}\}$ such that $|\mathbf{X}^{(d)}|/k_x^{(d)} \xrightarrow{D} R$ where $R$ has no point mass at $\mathbf{0}$ then for fixed $\mu$:*

(i) *If there exist $\{k_y^{(d)}\}$ such that $|\mathbf{Y}^{(d)}|/k_y^{(d)} \xrightarrow{D} Y$ then*

$$\overline{\alpha}_d(\mu) \to 2 \mathbb{E}\left[ \Phi\left( -\frac{\mu Y}{R} \right) \right]. \tag{13}$$

(ii) *If in fact $|\mathbf{Y}^{(d)}|/k_y^{(d)} \xrightarrow{m.s.} Y$ with $\mathbb{E}[Y^2] < \infty$ then*

$$\frac{d}{4 k_x^{(d)^2}} S_d^2(\mu) \to 2\mu^2 \mathbb{E}\left[ Y^2 \Phi\left( -\frac{\mu Y}{R} \right) \right]. \tag{14}$$

The remainder of this paper focusses on an important corollary to Theorem 3, which is obtained by setting $Y = 1$.

**Corollary 3.** *Let* $\{\mathbf{X}^{(d)}\}$, $\{\mathbf{Y}^{(d)}\}$, $\{k_x^{(d)}\}$ *and* $\{k_y^{(d)}\}$ *be as defined in Theorem 3 and let* $R$ *be any non-negative random variable with no point mass at* 0.

(i) *If* $|\mathbf{X}^{(d)}|/k_x^{(d)} \xrightarrow{D} R$ *and* $|\mathbf{Y}^{(d)}|/k_y^{(d)} \xrightarrow{m.s.} 1$

$$\overline{\alpha}_d(\mu) \to 2\mathbb{E}\left[\Phi\left(-\frac{\mu}{R}\right)\right], \tag{15}$$

$$\frac{d}{4k_x^{(d)^2}} S_d^2(\mu) \to 2\mu^2 \mathbb{E}\left[\Phi\left(-\frac{\mu}{R}\right)\right]. \tag{16}$$

(ii) *If* $|\mathbf{X}^{(d)}|/k_x^{(d)} \xrightarrow{p} 1$ *and* $|\mathbf{Y}^{(d)}|/k_y^{(d)} \xrightarrow{m.s.} 1$

$$\overline{\alpha}_d(\mu) \to 2\Phi(-\mu), \tag{17}$$

$$\frac{d}{4k_x^{(d)^2}} S_d^2(\mu) \to 2\mu^2 \Phi(-\mu). \tag{18}$$

With these asymptotic forms for EAR and ESJD we are finally equipped to examine the issue of optimal scaling in the limit as $d \to \infty$.

## 3.2. The validity and existence of an asymptotically optimal scaling

It was shown in Section 2 that there is at least one finite optimal scaling for any spherically symmetric unimodal finite dimensional target with finite second moment provided the second moment of the proposal is also finite. We now investigate the validity and existence of a finite asymptotically optimal (transformed) scaling for spherically symmetric targets as $d \to \infty$.

1. **Validity:** We shall obtain an asymptotically optimal scaling by maximising the limiting efficiency function. Ideally we would instead find the limit of the sequence of scalings which maximise each finite dimensional efficiency function. We investigate the circumstances under which these are equivalent.
2. **Existence:** It is not always the case that the limiting efficiency function possesses a finite maximum; examples are provided.

An even stronger validity assumption is implicit in works such as Roberts *et al.* [9], Roberts and Rosenthal [10] and Bedard [2]. In each of these papers a limiting process is found and the efficiency of this limiting process is maximised to give an asymptotically optimal scaling.

For a given sequence of targets and proposals with optimal scalings $\hat{\lambda}_d$, we seek the limiting transformed optimal scaling $\hat{\mu} := \lim_{d \to \infty} \hat{\mu}_d$, where $\hat{\mu}_d$ is given in terms of $\hat{\lambda}_d$ by (12). The optimal scaling as $d \to \infty$ would therefore be $\hat{\lambda}_d \sim (2k_x^{(d)}\hat{\mu})/(d^{1/2}k_y^{(d)})$. However the value $\hat{\mu}$ will be obtained by maximising $2\mu^2\Theta(-\mu) \propto \lim_{d\to\infty} S_d^2(\mu)$, where $\Theta$ is defined as in Theorem 2. The following result indicates when the scaling that optimises the limit is equivalent to the limit of the optimal scalings. A proof is provided in Appendix A.3.

**Lemma 2.** *Let $\{S_d^2(\mu)\}$ be a sequence of functions defined on $[0,\infty)$ with continuous pointwise limit $S^2(\mu)$. Define*

$$M := \left\{\arg\max_\mu S^2(\mu)\right\} \quad and \quad M_d := \left\{\arg\max_\mu S_d^2(\mu)\right\}.$$

*For each $d \in \mathbb{N}$ select any $\hat{\mu}_d \in M_d$.*

(i) *If $M = \{\hat{\mu}\}$ and $\hat{\mu}_d < a < \infty$ $\forall d$ then $\hat{\mu}_d \to \hat{\mu}$.*
(ii) *If $M = \{\hat{\mu}\}$ $\exists$ a sequence $\mu_d^* \to \hat{\mu}$ where each $\mu_d^*$ is a local maximum of $S_d^2(\cdot)$.*
(iii) *If $M = \phi$ ($S^2$ has no finite maximum) then $\hat{\mu}_d \to \infty$, that is, $\lim_{d\to\infty}(\min M_d) = \infty$.*
(iv) *If $M \neq \phi$ and $\hat{\mu}_d < a < \infty$ $\forall d$ then $S_d^2(\hat{\mu}_d) \to S^2(\hat{\mu})$ for any $\hat{\mu} \in M$.*

We now highlight certain aspects of Lemma 2 through reference to the mixture target (11), and specifically to target examples 6–8 from Section 2.1. Later in this section Lemma 2 is also applied to target examples 1–5. In all that follows consider the sequence of graphs of $S_d^2(\mu_d)$ against $\mu_d$, where $\mu_d$ is given by (12); consider also the graph of the pointwise limit, $S^2(\mu)$, against $\mu$. For all targets of the form (11) with sufficiently large $d$ (so that the components are sufficiently separated in scale) each graph of $S_d^2(\mu_d)$ against $\mu_d$ has two peaks, and a different rescaling $k_x^{(d)}$ applies to each component of the mixture. Choosing to rescale by the higher $k_x^{(d)}$ would stabilise the right-hand peak while the left would approach $\mu_d = 0$. However (unless $p_d \to 1$) this choice of scaling would create a point mass at the origin in the limiting rescaled radial distribution function $\overline{\Theta}(\cdot)$, which is forbidden in the statement of Theorem 3. In order to apply the theorem we must therefore rescale by the lower $k_x^{(d)}$ which stabilises the left-hand peak while the right-hand peak drifts off to $\mu_d = \infty$ *and is therefore not present in the pointwise limit*. The existence and consistency of a limiting optimal scaling then depend on the relative heights of the peaks which in turn depend on the limiting behaviour of $p_d$.

First consider any target with $p_d > 1/d^2$ such as target examples 6 and 7. For a given dimension this would produce plots similar to the right-hand panels of Figure 1 but with the right-hand peak higher than the left-hand peak and therefore providing the optimal scaling, $\hat{\mu}_d$. The limit of the scalings which maximise each finite dimensional ESJD is therefore not the same as the scaling which optimises the limiting ESJD. In Lemma 2 Parts (i) and (iv) this situation is prevented through the condition $\hat{\mu}_d < a < \infty$.

Suppose in fact that $p_d \to p > 0$, so that rescaling via the lower $k_x^{(d)}$ produces a point mass at $\infty$ in the limiting rescaled radial distribution. Consider first the optimal scaling obtained from the limiting form of the ESJD. By Theorem 2, $\Theta(-x_1) \to p/2$ as $x_1 \to \infty$. Hence the limiting ESJD given in Corollary 3 increases without bound as $\mu \to \infty$; this is an example of case (iii) in Lemma 2. The optimal scaling for exploring a real $d$-dimensional target follows the portion of the target with the larger scale parameter, and so in the limit accepted jumps only arise from this portion of the target. The limit of the optimal EAR is therefore the limiting optimal EAR for the larger component multiplied by a factor $p$, as suggested by the results for target example 6.

If $p_d \to p = 0$ then only the left-hand peak affects the forms in Corollary 3; the optimal scaling and acceptance rate calculated from this corollary are therefore identical to those for target example 1. However the true optimal scaling follows the right-hand peak and so the true limiting optimal acceptance rate is 0, as suggested by the results for target example 7.

Alternatively if $p_d < 1/d^2$ then for large enough $d$ the stabilised left-hand peak dominates, $\hat{\mu}_d$ is bounded and the limit of the maxima is the maximum of the limit function. The true limiting optimal acceptance rate is exactly that of the lower component as suggested for target example 8, and this is given correctly by Corollary 3.

Provided $p_d \to 0$ the limiting forms for EAR and ESJD are unaffected by the speed at which this limit is approached. The limiting forms are therefore uninformative about whether or not the second peak is important. This is a fundamental issue with the identifiability of a limiting optimal scaling from the limiting ESJD.

The above clearly generalises from the specific form (11) so that failure of the boundedness condition on $\hat{\mu}_d$ in Lemma 2 intuitively corresponds to a *target sequence that contains a mixture of scales that produce local maxima in ESJD and whose ratio increases without bound*. In general, targets that vary on at least two very different scales are not amenable to the current approach. Indeed the very existence of a single "optimal" scaling is highly debatable. We wish to work with the limit $S^2(\mu)$, accepting its potential limitation. Therefore define $\hat{\mu} := \min M$ (or $\hat{\mu} := \infty$ if $M = \phi$), to be the **asymptotically optimal transformed scaling** (AOTS), and $\hat{\lambda}_d = (2k_x^{(d)}\hat{\mu})/(d^{1/2}k_y^{(d)})$ to be the **asymptotically optimal scaling** (AOS). These are equivalent to the limit of the optimal (transformed) scalings provided $\hat{\mu}_d < a < \infty, \forall d$. Similarly the **asymptotically optimal expected acceptance rate** (AOA) is the limiting EAR that results from using the AOTS.

We now turn to the existence of an asymptotically optimal scaling. The practising statistician is free to choose the proposal distribution and we therefore assume throughout the remainder of our discussion of spherically symmetric targets that there is a sequence $k_y^{(d)}$ such that the transformed proposal satisfies $|\mathbf{Y}^{(d)}|/k_y^{(d)} \xrightarrow{m.s.} 1$.

First consider the special case where there is a sequence $k_x^{(d)}$ such that the transformed target satisfies $|\mathbf{X}^{(d)}|/k_x^{(d)} \xrightarrow{p} 1$. Differentiating (18) we see that the optimal scaling must satisfy $2\Phi(-\hat{\mu}_p) = \hat{\mu}_p\phi(-\hat{\mu}_p)$, which gives $\hat{\mu}_p :\approx 1.19$. Substituting into (17) provides the EAR at this optimal scaling: $\hat{\alpha}_p :\approx 0.234$, as suggested by the finite dimensional results for target examples 1 and 2. More generally $|\mathbf{X}^{(d)}|/k_x^{(d)} \xrightarrow{D} R$. Following our discussions on validity we now assume that $R$ contains no point mass at 0 or $\infty$. In general we seek a finite scaling $\hat{\mu}$ that maximises the pointwise limit of $S_d^2$ as given in (16); we then compute the EAR using (15). We illustrate this process with reference to three of our non-standard examples from Section 2.1.

For target examples 3 and 4 the marginal radial distribution is positive Gaussian ($\overline{\theta}_d(r) \propto e^{-r^2/2}$) and exponential ($\overline{\theta}_d(r) = e^{-r}$), respectively. $S^2(\mu)$ is maximised at $\hat{\mu} =$ 1.67 and $\hat{\mu} = 2.86$, respectively, which correspond to EARs of 0.091 and 0.055, consistent with the findings in Section 2.1. In target examples 1–4, the ESJD of each element in the sequence has a single maximum, so by Lemma 2(ii) the limit of these maxima is the maximum of the limit function, subject to the scaling $k_x^{(d)}$.

For target example 5 the limiting transformed marginal radial density is $\overline{\theta}(r) \propto e^{-(\log r)^2}$ and numerical evaluation shows $S^2(\mu)$ to be bounded above but to increase monotonically with $\mu$; this corresponds to case (iii) of Lemma 2. As with target example 6 this provides a situation where $S^2(\mu)$ is increasing as $\mu \to \infty$. However unlike target example 6, here there is no radial point mass at $\infty$, $S^2(\mu)$ is bounded and the limiting optimal EAR is 0.

### 3.3. Asymptotically optimal scaling and EAR

If $|\mathbf{Y}^{(d)}|/k_y^{(d)} \xrightarrow{m.s.} 1$ then for $\mu$ to be optimal we require

$$2\Theta(-\mu) = \mu\Theta'(-\mu). \tag{19}$$

There may not always be a solution for $\mu$ (see Section 3.2) but when there is, denote this value as $\hat{\mu}$. Asymptotically optimal scaling is therefore achieved by setting $\mu = \hat{\mu}$, so that rearranging (12) we obtain the following corollary to Theorem 3:

**Corollary 4.** *Let $\{\mathbf{X}^{(d)}\}$ be a sequence of $d$-dimensional spherically symmetric unimodal target distributions and let $\{\mathbf{Y}^{(d)}\}$ be a sequence of jump proposal distributions. If there exist sequences $\{k_x^{(d)}\}$ and $\{k_y^{(d)}\}$ such that the marginal radial distribution function of $\mathbf{X}^{(d)}$ satisfies $|\mathbf{X}^{(d)}|/k_d \xrightarrow{D} R$ where $R$ has no point mass at $\mathbf{0}$, $|\mathbf{Y}^{(d)}|/k_y^{(d)} \xrightarrow{m.s.} 1$, and provided there is a solution $\hat{\mu}$ to (19) then the asymptotically optimal scaling (AOS) satisfies*

$$\hat{\lambda}_d = 2\hat{\mu}\frac{k_x^{(d)}}{d^{1/2}k_y^{(d)}}.$$

Target examples 1 and 2 satisfy $|\mathbf{X}^{(d)}|/k_x^{(d)} \xrightarrow{p} 1$, with $k_x^{(d)} = d^{1/2}$ and $k_x^{(d)} = d$ respectively; we therefore expect an AOTS of $\hat{\mu}_p \approx 1.19$. For target examples 3 and 4, $k_x^{(d)} = 1$, and the AOTSs are, respectively, $\hat{\mu} \approx 1.67$ and $\hat{\mu} \approx 2.86$. Figure 4 shows plot of the true optimal scale parameter against dimension evaluated numerically using (9). The asymptotic approximation given in Corollary 4 appears as a dotted line. Both axes are log-transformed and in all cases the finite dimensional optimal scalings are seen to approach their asymptotic values as $d$ increases. For the Gaussian target very close agreement is attained even in one dimension since the asymptotic Gaussian approximation to the marginal radial distribution function is exact for all finite $d$.

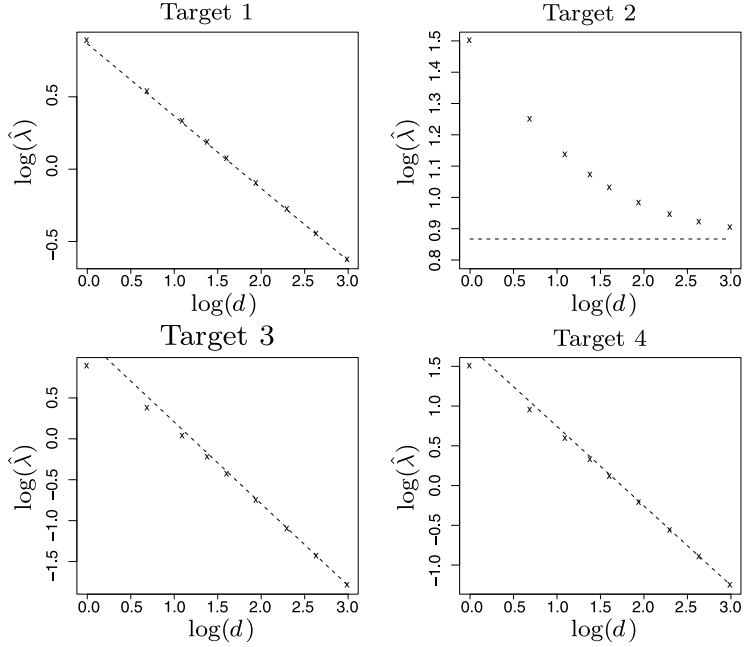

**Figure 4.** Plots of $\log \hat{\lambda}$ against $\log d$ for target examples 1–4. Optimal values predicted using Corollary 4 appear as dotted lines.

Let us now explore the AOA, if it exists, and define $\overline{\alpha}_\infty(\mu) := \lim_{d \to \infty} \overline{\alpha}_d(\mu)$. From Theorem 2 and Corollary 3(i)

$$\overline{\alpha}_\infty(\mu) = 2\Theta(-\mu) = 2\mathbb{E}_R\left[\Phi\left(\frac{-\mu}{R}\right)\right],$$

where $R$ is the marginal radius of the limit of the sequence of scaled targets. We build upon Theorem 3, which explicitly requires that the rescaled marginal radial distribution should have no point mass at 0. Following the discussion in Section 3.2 the condition that there be an optimal $\mu$ implies that the limiting marginal radius has no point mass at infinity. The following is proved in Appendix A.4.

**Theorem 4.** *Let $\mathbf{X}^{(d)}$ be a sequence of $d$-dimensional spherically symmetric unimodal target distributions and let $\mathbf{Y}^{(d)}$ be a sequence of jump proposal distributions. Let there exist $k_x^{(d)}$ and $k_y^{(d)}$ such that $|\mathbf{Y}^{(d)}|/k_y^{(d)} \overset{m.s.}{\longrightarrow} 1$ and $|\mathbf{X}^{(d)}|/k_x^{(d)} \overset{D}{\longrightarrow} R$ for some $R$ with no point mass at 0. If there is a limiting (non-zero) AOA it is*

$$\overline{\alpha}_\infty(\mu) \leq \hat{\alpha}_p \approx 0.234.$$

*Equality is achieved if and only if there exist $k_x^{(d)}$ such that $|\mathbf{X}^{(d)}|/k_x^{(d)} \overset{p}{\longrightarrow} 1$.*

For "standard" proposals, the often used optimal EAR of 0.234 therefore provides an upper bound on the possible optimal EARs for spherically symmetric targets, and it is achieved if and only if the mass of the target converges (after rescaling) to an infinitesimally thin shell.

## 4. Elliptically symmetric distributions

As discussed in Section 1.2 a unimodal elliptically symmetric target $\mathbf{X}$ may be defined in terms of an associated orthogonal linear map $\mathbf{T}$ such that $\mathbf{X}_* := \mathbf{T}(\mathbf{X})$ is spherically symmetric with unit scale parameter. Since $\mathbf{T}$ is linear, the jump proposal in the transformed space is $\mathbf{Y}_* := T(\mathbf{Y})$.

The ESJD (4) is preserved under the transformation, since $Y_i^2/\beta_i^2 = Y_{i*}^2$, and thus we simply apply Theorem 1 in the transformed space. Write $F_{1|d}^*(\cdot)$ for the one-dimensional marginal density of spherically symmetric $\mathbf{X}_*$, We wish to optimise the ESJD

$$S_d^2(\lambda) := 2\lambda^2 \mathbb{E}[|\mathbf{Y}_*|^2 F_{1|d}^*(-\tfrac{1}{2}\lambda|\mathbf{Y}_*|)]. \tag{20}$$

Here expectation is with respect to Lebesgue measure $r_*(\cdot)$ of $\mathbf{Y}_*$. Acceptance in the original space is equivalent to acceptance in the transformed space and the EAR is therefore given by

$$\overline{\alpha}_d(\lambda) = 2\mathbb{E}[F_{1|d}^*(-\tfrac{1}{2}\lambda|\mathbf{Y}_*|)]. \tag{21}$$

Corollaries 1 and 2 are now seen to hold for all unimodal elliptically symmetric targets. For Corollary 3(i) to be applicable in the transformed space we require there to exist $k_x^{*(d)}$ and $k_y^{*(d)}$ such that

$$\frac{|\mathbf{T}^{(d)}(\mathbf{X}^{(d)})|}{k_x^{*(d)}} \xrightarrow{D} R \quad \text{and} \quad \frac{|\mathbf{T}^{(d)}(\mathbf{Y}^{(d)})|}{k_y^{*(d)}} \xrightarrow{m.s.} 1. \tag{22}$$

Since $\mathbf{X}_*^{(d)}$ is spherically symmetric, it is natural to request convergence of $\mathbf{X}_*^{(d)}/k_x^{*(d)}$ explicitly in the statement of the theorem. Bedard [2] considers a situation analogous to this, but with $R = 1$. The working statistician is free to choose a jump proposal such that $|\mathbf{Y}^{(d)}|/k_y^{(d)} \xrightarrow{m.s.} 1$. If $\mathbf{Y}^{(d)}$ is in fact spherically symmetric this convergence carries through to the transformed space provided the eccentricity of the original target is not "too severe". A proof of the following appears in Appendix A.5.

**Theorem 5.** *Let $\{\mathbf{X}^{(d)}\}$ be a sequence of elliptically symmetric unimodal targets and $\{\mathbf{T}^{(d)}\}$ be a sequence of linear maps such that $\mathbf{X}_*^{(d)} := \mathbf{T}^{(d)}(\mathbf{X}^{(d)})$ is spherically symmetric with unit scale parameter. Let $\{\mathbf{Y}^{(d)}\}$ be a sequence of spherically symmetric proposals and let there exist $\{k_x^{*(d)}\}$ and $\{k_y^{(d)}\}$ such that*

$$\frac{\mathbf{X}_*^{(d)}}{k_x^{*(d)}} \xrightarrow{D} R \quad and \quad \frac{\mathbf{Y}^{(d)}}{k_y^{(d)}} \xrightarrow{m.s.} 1.$$

Denote by $\nu_i$ the eigenvalues of $\mathbf{T}^{(d)}$, and define $k_y^{*(d)} = (\overline{\nu^2})^{1/2} k_y^{(d)}$, where $\overline{\nu^2} := d^{-1} \sum_{i=1}^{d} \nu_i^2$. If

$$\frac{\nu_{\max}(d)^2}{\sum_{i=1}^{d} \nu_i(d)^2} \to 0 \tag{23}$$

then for fixed

$$\mu := \frac{1}{2} \frac{d^{1/2} k_y^{*(d)}}{k_x^{*(d)}} \lambda_d \tag{24}$$

the EAR and the ESJD satisfy

$$\overline{\alpha}_d(\mu) \to 2\mathbb{E}\left[\Phi\left(-\frac{\mu}{R}\right)\right], \tag{25}$$

$$\frac{d}{4 k_x^{*(d)2}} S_d^2(\mu) \to 2\mu^2 \mathbb{E}\left[\Phi\left(-\frac{\mu}{R}\right)\right], \tag{26}$$

where $\Phi(x)$ is the cumulative distribution function of a standard Gaussian. If in fact $R = 1$ then

$$\overline{\alpha}_d(\mu) \to 2\Phi(-\mu), \tag{27}$$

$$\frac{d}{4 k_x^{*(d)2}} S_d^2(\mu) \to 2\mu^2 \Phi(-\mu). \tag{28}$$

Naturally (28) leads to the same optimal $\hat{\mu}_p$ as for a spherically symmetric target, so the AOA is still approximately 0.234 and the AOS satisfies

$$\hat{\lambda}_d = 2\hat{\mu}_p \frac{k_x^{*(d)}}{d^{1/2} k_y^{(d)}} \times \frac{1}{(\overline{\nu^2})^{1/2}}.$$

Similarly (25) and (26) lead again to $\alpha(\hat{\mu}) \leq \alpha(\hat{\mu}_p) \approx 0.234$.

## 5. Discussion

We have investigated optimal scaling of the random walk Metropolis algorithm on unimodal elliptically symmetric targets. An approach through finite dimensions using expected square jumping distance (ESJD) as a measure of efficiency both agrees with and extends the existing literature, which is based upon diffusion limits.

We obtained exact analytical expressions for the expected acceptance rate (EAR) and the ESJD in finite dimension $d$. For any RWM algorithm on a spherically symmetric unimodal target it was shown that EAR decreases monotonically from 1 to 0 as the proposal scaling parameter increases from 0 to $\infty$. This bijective mapping justifies to an extent the use of acceptance rate as a proxy for the scale parameter. The theory for finite dimensional targets was then shown to extend to elliptically symmetric targets.

An asymptotic theory was developed for the behaviour of the RWM algorithm as dimension $d \to \infty$. It was shown that the asymptotically optimal EAR of 0.234 extends to the class of spherically symmetric unimodal targets if and only if the mass of the targets converges to a spherical shell that becomes infinitely thin relative to its radius, with a similar but slightly stronger condition on the proposal. The optimal acceptance rate was then explored for target sequences for which the "shell" condition fails. In such cases the asymptotically optimal EAR (if it exists) was shown to be strictly less than 0.234. An asymptotic form for the optimal scale parameter showed that the dimension dependent rescalings which stabilise the radial mass for both the proposal and target must be taken into account. Much of the existing literature (see Roberts *et al.* [9]) uses independent and identically distributed (i.i.d.) target components and i.i.d. proposal components so that these two extra effects cancel and $\hat{\lambda}_d \propto d^{-1/2}$.

The class for which the limit results are valid was then extended to include all algorithms on elliptically symmetric targets such that the same "shell" conditions are satisfied once the target has been transformed to spherical symmetry by an orthogonal linear map. If the original target is explored by a spherically symmetric proposal then an additional constraint applies to the eigenvalues of the linear map, which forbids the scale parameter of the smallest principle component from being "too much smaller" than all the other scale parameters and is equivalent to the condition of Bedard [2], derived for targets with independent components that are identical up to a scaling.

The optimality limit results are not always valid for targets with at least two very different (but important) scales of variation; however, the suitability of the RWM to such targets is itself questionable.

Explicit forms for EAR and ESJD in terms of marginal radial densities were also used to explore specific combinations of target and proposal in finite dimensions. Numerical and analytical results agreed with our limit theory and with a simulation study in Roberts and Rosenthal [10].

# Appendix

## A.1. Proof of Theorem 1

The proof of Theorem 1 relies on a partitioning of the space of possible values for $\mathbf{x}^*$ (and so for $\mathbf{x}'$) given $\mathbf{x}$ into four disjoint regions:

- the **identity** region: $R_{\mathrm{id}}(\mathbf{x}) := \{\mathbf{x}\}$,
- the **equality** region: $R_{\mathrm{eq}}(\mathbf{x}) := \{\mathbf{x}' \in \Re^d : \mathbf{x}' \notin R_{\mathrm{id}}(\mathbf{x}), \frac{\pi(\mathbf{x}')q(\mathbf{x}|\mathbf{x}')}{\pi(\mathbf{x})q(\mathbf{x}'|\mathbf{x})} = 1\}$,
- the **acceptance** region: $R_{\mathrm{a}}(\mathbf{x}) := \{\mathbf{x}' \in \Re^d : \alpha(\mathbf{x}, \mathbf{x}') = 1, \mathbf{x}' \notin R_{\mathrm{eq}}(\mathbf{x}) \cup R_{\mathrm{id}}(\mathbf{x})\}$,
- the **rejection** region: $R_{\mathrm{r}}(\mathbf{x}) := \{\mathbf{x}' \in \Re^d : \alpha(\mathbf{x}, \mathbf{x}') < 1\}$.

Here $\pi(\cdot)$ is the target (Lebesgue) density. For vectors $(\mathbf{x}, \mathbf{x}')$ in $\Re^d \times \Re^d$ we employ the shorthand $R_{\mathrm{ID}} := \{(\mathbf{x}, \mathbf{x}') : \mathbf{x} \in \Re^d, \mathbf{x}' \in R_{\mathrm{id}}(\mathbf{x})\}$, with regions $R_{\mathrm{EQ}}, R_{\mathrm{A}}$ and $R_{\mathrm{R}}$ defined analogously.

The following lemma holds for almost any Metropolis–Hastings algorithm and allows us to simplify the calculations of ESJD and EAR. It is convenient to be able to refer to the proposed jump, $\mathbf{Y}^* := \mathbf{X}^* - \mathbf{X}$.

**Lemma 3.** *Consider any Metropoplis–Hastings Markov chain with stationary Lebesgue density $\pi(\cdot)$. At stationarity let $\mathbf{X}$ denote the current element, $\mathbf{X}^*$ the proposed next element and $\mathbf{X}'$ the realised next element. Let proposals be drawn from Lebesgue density $q(\mathbf{x}^*|\mathbf{x})$ and assume that*

$$\int_{R_{\text{eq}}(\mathbf{x})} d\mathbf{x}' q(\mathbf{x}'|\mathbf{x}) = 0 \qquad \forall \mathbf{x}. \tag{29}$$

*Also denote the probability of accepting proposal $\mathbf{x}^*$ by $\alpha(\mathbf{x}, \mathbf{x}^*)$ and the joint laws of $(\mathbf{X}, \mathbf{X}^*)$ and $(\mathbf{X}, \mathbf{X}')$, respectively, by*

$$A^*(d\mathbf{x}, d\mathbf{x}^*) := \pi(\mathbf{x}) \, d\mathbf{x} \, q(\mathbf{x}^*|\mathbf{x}) \, d\mathbf{x}^* \tag{30}$$

*and*

$$A(d\mathbf{x}, d\mathbf{x}') := A^*(d\mathbf{x}, d\mathbf{x}')\alpha(\mathbf{x}, \mathbf{x}') \mathbb{1}_{\{\mathbf{x}' \neq \mathbf{x}\}}$$
$$+ \pi(\mathbf{x}) \, d\mathbf{x} \int d\mathbf{x}^* q(\mathbf{x}^*|\mathbf{x})(1 - \alpha(\mathbf{x}, \mathbf{x}^*)) \mathbb{1}_{\{\mathbf{x}' = \mathbf{x}\}}. \tag{31}$$

*Finally let $h(\mathbf{x}, \mathbf{x}')$ be any function satisfying the following two conditions:*

$$h(\mathbf{x}, \mathbf{x}') = c \times h(\mathbf{x}', \mathbf{x}) \qquad \forall \mathbf{x}, \mathbf{x}' \text{ (with } c = \pm 1), \tag{32}$$

$$h(\mathbf{x}, \mathbf{x}) = 0 \qquad \forall \mathbf{x}. \tag{33}$$

*Subject to the above conditions:*

1. $\mathbb{E}[h(\mathbf{X}, \mathbf{X}')] = (1 + c) \times \int_{(\mathbf{x}, \mathbf{x}') \in R_A} d\mathbf{x} \, d\mathbf{x}' \, \pi(\mathbf{x}) q(\mathbf{x}'|\mathbf{x}) h(\mathbf{x}, \mathbf{x}')$.
2. $\mathbb{E}[\alpha(\mathbf{X}, \mathbf{X}^*)] = 2 \int_{(\mathbf{x}, \mathbf{x}^*) \in R_A} d\mathbf{x} \, d\mathbf{x}^* \, \pi(\mathbf{x}) q(\mathbf{x}^*|\mathbf{x})$.

**Proof.** First note that an exchangeability between the regions $R_a(\cdot)$ and $R_r(\cdot)$ follows directly from their definitions

$$\mathbf{x}' \in R_a(\mathbf{x}) \quad \Longleftrightarrow \quad \mathbf{x} \in R_r(\mathbf{x}').$$

Consecutively applying this exchangeability, reversibility and the symmetry of $h(\cdot, \cdot)$, we find:

$$\int_{(\mathbf{x}, \mathbf{x}') \in R_A} A(d\mathbf{x}, d\mathbf{x}') h(\mathbf{x}, \mathbf{x}') = \int_{(\mathbf{x}', \mathbf{x}) \in R_R} A(d\mathbf{x}, d\mathbf{x}') h(\mathbf{x}, \mathbf{x}')$$
$$= \int_{(\mathbf{x}', \mathbf{x}) \in R_R} A(d\mathbf{x}', d\mathbf{x}) h(\mathbf{x}, \mathbf{x}')$$

$$= c \times \int_{(\mathbf{x}',\mathbf{x}) \in R_R} A(\mathrm{d}\mathbf{x}', \mathrm{d}\mathbf{x}) h(\mathbf{x}', \mathbf{x}).$$

The set $R_{\mathrm{ID}}$ corresponds to the second term in $A(\mathrm{d}\mathbf{x}, \mathrm{d}\mathbf{x}')$; this is in general not null with respect to $A(\cdot, \cdot)$; however, (33) implies that $h(\mathbf{x}, \mathbf{x}') = 0$ in $R_{\mathrm{ID}}$. Further, $\alpha(\mathbf{x}, \mathbf{x}^*) = 1 \; \forall (\mathbf{x}, \mathbf{x}^*) \in R_{\mathrm{EQ}}(\mathbf{x}, \mathbf{x}^*)$, and (29) holds. Therefore

$$\int_{(\mathbf{x},\mathbf{x}') \in R_{\mathrm{ID}} \cup R_{\mathrm{EQ}}} A(\mathrm{d}\mathbf{x}, \mathrm{d}\mathbf{x}') h(\mathbf{x}, \mathbf{x}') = 0.$$

Thus $R_{\mathrm{ID}}$ and $R_{\mathrm{EQ}}$ contribute nothing to the overall expectation of $h(\mathbf{X}, \mathbf{X}')$. Since $\alpha(\mathbf{x}, \mathbf{x}^*) = 1 \; \forall (\mathbf{x}, \mathbf{x}^*) \in R_A(\mathbf{x}, \mathbf{x}^*)$ the first result then follows.

The proof of the second result is similar to that of the first and is omitted. □

Sherlock [11] shows that for a symmetric proposal (such as the RWM) Lemma 3 may be extended to deal with cases where the density contains a series of plateaux and hence $R_{\mathrm{EQ}}$ is not null. $R_{\mathrm{EQ}}$ is then partitioned into a null set and pseudo-acceptance and rejection regions that are exactly as would be found if each plateau in fact had a small downward slope away from the origin.

In the region $R_A$, where acceptance is guaranteed, we have $\mathbf{x}' = \mathbf{x}^*$ and $\mathbf{y} = \mathbf{y}^*$ so that for integrals over $R_A$ we need not distinguish between proposed and accepted values. Applying Lemma 3 with $h(\mathbf{x}, \mathbf{x}') = \|\mathbf{x}' - \mathbf{x}\|_\beta^2 = |\mathbf{y}|^2$, we have

$$\overline{\alpha}_d(\lambda) = \frac{2}{\lambda^d} \int_{R_A} \mathrm{d}\mathbf{x} \, \mathrm{d}\mathbf{y} \, \pi(\mathbf{x}) r(\mathbf{y}/\lambda), \tag{34}$$

$$S_d^2(\lambda) = \frac{2}{\lambda^d} \int_{R_A} \mathrm{d}\mathbf{x} \, \mathrm{d}\mathbf{y} \, |\mathbf{y}|^2 \pi(\mathbf{x}) r(\mathbf{y}/\lambda). \tag{35}$$

First consider target densities that decrease with strict monotonicity from the mode. In this case $R_A$ corresponds to the region where

$$\pi(\mathbf{x} + \mathbf{y}) > \pi(\mathbf{x}) \iff |\mathbf{x} + \mathbf{y}|^2 < |\mathbf{x}|^2 \iff \mathbf{x} \cdot \hat{\mathbf{y}} < -\tfrac{1}{2}|\mathbf{y}|, \tag{36}$$

where $\hat{\mathbf{y}}$ is the unit vector in the direction of $\mathbf{y}$. So

$$(\mathbf{x}, \mathbf{x} + \mathbf{y}) \in R_A \iff \mathbf{y} \in \Re^d \quad \text{and} \quad \mathbf{x} \cdot \hat{\mathbf{y}} < -\tfrac{1}{2}|\mathbf{y}|.$$

Thus (34) and (35) become

$$\overline{\alpha}_d(\lambda) = \frac{2}{\lambda^d} \int_{\Re^d} \mathrm{d}\mathbf{y} \, r(\mathbf{y}/\lambda) F_{1|d}\left(-\frac{1}{2}|\mathbf{y}|\right),$$

$$S_d^2(\lambda) = \frac{2}{\lambda^d} \int_{\Re^d} \mathrm{d}\mathbf{y} \, |\mathbf{y}|^2 r(\mathbf{y}/\lambda) F_{1|d}\left(-\frac{1}{2}|\mathbf{y}|\right),$$

and the required results follow directly.

Without strict monotonicity $R_A$ must simply be extended to include the "pseudo-acceptance regions" defined in Sherlock [11].

### A.2. Proof of Theorem 3

Theorem 3 follows almost directly from the following simple lemma, the proof of which follows from a standard measure theory argument that we omit.

**Lemma 4.** *Let $U_d$ be a sequence of random variables and let $G_d(\cdot) \to G(\cdot)$ be a sequence of monotonic functions with $0 \leq G_d(u) \leq 1$ and $G(\cdot)$ continuous. Then*

$$U_d \xrightarrow{p} U \quad \Longrightarrow \quad \mathbb{E}[G_d(U_d)] \to \mathbb{E}[G(U)], \quad and$$

$$U_d \xrightarrow{m.s.} U \quad \Longrightarrow \quad \mathbb{E}[U_d^2 G_d(U_d)] \to \mathbb{E}[U^2 G(U)].$$

Now note that $-\frac{1}{2}\lambda_d |\mathbf{Y}^{(d)}| = -\mu |\mathbf{Y}^{(d)}|/k_y^{(d)} \times k_x^{(d)}/d^{1/2}$ whence (5) and (6) become

$$\overline{\alpha}_d(\mu) = 2\mathbb{E}\left[F_{1|d}\left(-\mu \frac{|\mathbf{Y}^{(d)}|}{k_y^{(d)}} \frac{k_x^{(d)}}{d^{1/2}}\right)\right] \quad \text{and}$$

$$S_d^2(\mu) = \frac{8\mu^2 k_x^{(d)2}}{d} \mathbb{E}\left[\left(\frac{|\mathbf{Y}^{(d)}|}{k_y^{(d)}}\right)^2 F_{1|d}\left(-\mu \frac{|\mathbf{Y}^{(d)}|}{k_y^{(d)}} \frac{k_x^{(d)}}{d^{1/2}}\right)\right].$$

Again denote the limiting one-dimensional distribution function corresponding to $R$ by $\Theta(x)$. In Lemma 4 substitute $U_d = |\mathbf{Y}^{(d)}|/k_y^{(d)}$, $U = Y$, $G_d(u) = F_{1|d}(-\mu \frac{k_x^{(d)}}{d^{1/2}} u)$ and $G(u) = \Theta(-\mu u)$. Note also that since $G(\cdot)$ and $G_d(\cdot)$ are bounded the convergence in the first part of the lemma holds if $U_d \xrightarrow{D} U$. The theorem then follows directly.

### A.3. Proof of Lemma 2

(i) Pick an arbitrarily small $\delta > 0$ and set $a^* = \max(a, \hat{\mu} + \delta)$.
Define $R := (\hat{\mu} - \delta, \hat{\mu} + \delta)$ and $T := [0, a^*] \backslash R$.
Let $m := \max_{\mu \in T} S^2(\mu)$. Since $\hat{\mu} \in R$ uniquely maximises $S^2(\cdot)$ in $[0, a^*]$, and since $T$ is compact, a strict inequality holds: $m < S^2(\hat{\mu})$.
Also $S_d^2(\mu) \to S^2(\mu)$ uniformly on compact $[0, a^*]$ and hence $\exists d_1$ such that

$$|S^2(\mu) - S_d^2(\mu)| < \tfrac{1}{2}(S^2(\hat{\mu}) - m) \qquad \forall \mu \in [0, a^*] \text{ and } d > d_1.$$

Hence for any $\mu_b \in T$ and $d > d_1$

$$S_d^2(\mu_b) < S^2(\mu_b) + \tfrac{1}{2}(S^2(\hat{\mu}) - m) \leq \tfrac{1}{2}(S^2(\hat{\mu}) + m)$$
$$= S^2(\hat{\mu}) - \tfrac{1}{2}(S^2(\hat{\mu}) - m) < S_d^2(\hat{\mu}).$$

Since $\hat{\mu}_d$ is confined to $[0, a^*]$ it must therefore reside in $R$.

(ii) Pick an arbitrarily small $\delta > 0$ and set $a^* = \hat{\mu} + 2\delta$. Proceed exactly as in the proof to (i). The interval $(\hat{\mu} - \delta, \hat{\mu} + \delta)$ contains a local maximum of $S_d^2$ since $S_d^2(\mu_b) < S_d^2(\hat{\mu})$ for $d > d_1$ and $\mu_b \in [0, a^*] \backslash (\hat{\mu} - \delta, \hat{\mu} + \delta)$.

(iii) Pick a large $k > 0$ and let $m := \max_{\mu \in [0,k]} S^2(\mu)$. Now $m < \infty$ since $S^2(\mu)$ is continuous on compact $[0,k]$. As there is no finite arg max, $\exists \mu_b > k$ such that $S^2(\mu_b) = m + \delta$ for some $\delta > 0$.

Convergence of $S_d^2(\mu)$ to $S^2(\mu)$ is uniform on $[0, \mu_b]$ and therefore $\exists d_1$ such that

$$|S^2(\mu) - S_d^2(\mu)| < \frac{\delta}{2} \qquad \forall \mu \in [0, \mu_b] \text{ and } d > d_1.$$

Hence for $\mu_a \in [0, k]$

$$S_d^2(\mu_a) < S^2(\mu_a) + \frac{\delta}{2} = S^2(\mu_b) - \frac{\delta}{2} < S_d^2(\mu_b).$$

Thus for any $k$ and any $\mu_a \in [0, k]$, for all large enough $d$ there is always an $\mu_b > k$ such that $S_d^2(\mu_b) > S_d^2(\mu_a)$ and hence the maximum is achieved at $\mu_d > k$. Therefore $\mu_d \to \infty$.

(iv) Define $a^* = \max(a, \hat{\mu})$ and choose an $\varepsilon > 0$. Since $S_d^2(\mu) \to S^2(\mu)$ uniformly on compact $[0, a^*]$, $\exists d_1$ such that $\forall d > d_1$ and $\mu \in [0, a^*]$, $|S^2(\mu) - S_d^2(\mu)| < \varepsilon$. By definition $S_d^2(\hat{\mu}_d) \geq S_d^2(\hat{\mu})$, therefore

$$S^2(\hat{\mu}) - \varepsilon < S_d^2(\hat{\mu}) \leq S_d^2(\hat{\mu}_d) < S^2(\hat{\mu}_d) + \varepsilon < S^2(\hat{\mu}) + \varepsilon.$$

## A.4. Proof of Theorem 4

Observe that

$$\Theta'(-\mu) = \mathbb{E}\left[\frac{1}{R}\phi\left(-\frac{\mu}{R}\right)\right],$$

so (19) becomes

$$2\mathbb{E}\left[\Phi\left(-\frac{\mu}{R}\right)\right] = \mathbb{E}\left[\frac{\mu}{R}\phi\left(-\frac{\mu}{R}\right)\right]. \tag{37}$$

For a given distribution of $R$, this has solution $\hat{\mu}$, from which the AOA is

$$\hat{\alpha} := \overline{\alpha}_\infty(\hat{\mu}) = 2\mathbb{E}\left[\Phi\left(-\frac{\hat{\mu}}{R}\right)\right].$$

Substitute $V := \Phi(-\frac{\hat{\mu}}{R})$ so that for $\mu \geq 0$ and $R \geq 0$ we have $v \in [0, 0.5]$. Also define

$$h(v) := -\Phi^{-1}(v)\phi(\Phi^{-1}(v)).$$

The AOA is therefore

$$\hat{\alpha} = 2\mathbb{E}[V]$$

and (37) is satisfied, becoming

$$2\mathbb{E}[V] = \mathbb{E}[h(V)]. \tag{38}$$

But

$$\frac{\mathrm{d}^2 h}{\mathrm{d}v^2} = 2\frac{\Phi^{-1}(v)}{\phi(\Phi^{-1}(v))} \leq 0 \quad \text{for} v \in [0, 0.5],$$

with strict inequality for $v \in (0, 0.5)$ (i.e., $r \in (0, \infty)$). Therefore by Jensen's inequality

$$\mathbb{E}[h(V)] \leq h(\mathbb{E}[V]). \tag{39}$$

Since $h''(\cdot)$ is strictly negative except at the (finite) end points, equality is achieved if and only if all the mass in $V$ is concentrated in one place, $v_0$; this corresponds to all the mass in $R$ being concentrated at $-\hat{\mu}/\Phi^{-1}(v_0)$ and is exactly the situation $|\mathbf{X}^{(d)}|/k_x^{(d)} \xrightarrow{p} 1$.

Substitute $m := \Phi^{-1}(\mathbb{E}[V])$, so that (38) and (39) combine to give

$$2\Phi(-m) \leq m\phi(m).$$

When there is equality the single solution to this equation is $\hat{m} = \hat{\mu}_p \approx 1.19$. The inequality is strict if and only if $m > \hat{\mu}_p$, and hence $2\Phi(-m) \leq 2\Phi(-\hat{\mu}_p)$.

Therefore the AOA is

$$\hat{\alpha} = \mathbb{E}[V] = 2\Phi(-m) \leq 2\phi(-\hat{\mu}_p) \approx 0.234,$$

with equality achieved if and only if $|\mathbf{X}^{(d)}|/k_x^{(d)} \xrightarrow{p} 1$.

## A.5. Proof of Theorem 5

Denote the arithmetic mean of the squares of the $d$-dependent scalar values $\alpha_1(d), \ldots, \alpha_d(d)$ by $\overline{\alpha^2}(d)$, and the maximum by $\alpha_{\max}(d)$. The following is proved in Sherlock [11].

**Lemma 5.** *Let $\mathbf{S}^{(d)}$ be a sequence of orthogonal linear maps on $\Re^d$ with eigenvalues $\alpha_1(d), \ldots, \alpha_d(d)$ and let $\mathbf{U}^{(d)}$ be a sequence of isotropic random variables in $\Re^d$. Then*

$$|\mathbf{U}^{(d)}| \xrightarrow{m.s.} 1 \quad \Longleftrightarrow \quad \frac{|\mathbf{S}^{(d)}(\mathbf{U}^{(d)})|}{(\overline{\alpha^2}(d))^{1/2}} \xrightarrow{m.s.} 1$$

*provided the eigenvalues of $\mathbf{S}^{(d)}$ satisfy*

$$\frac{\alpha_{\max}(d)^2}{\sum_{i=1}^d \alpha_i(d)^2} \to 0. \tag{40}$$

It should be emphasised that condition (40) applies to the map that transforms a spherically symmetric random variable ***to*** an elliptically symmetric random variable.

Now apply Lemma 5 with $\mathbf{U}^{(d)} = \mathbf{Y}^{(d)}/k_y^{(d)}$ and $\mathbf{S}^{(d)} = \mathbf{T}^{(d)}$ to see that the second half of (22) holds with the new rescaling factor $k_y^{*(d)}$. We may therefore apply Corollary 3 in the transformed space with $\mu$ as defined in (24). Since EAR and ESJD are invariant to the transformation this leads directly to (25)–(28).

# References

[1] Apostol, T.M. (1974). *Mathematical Analysis*. Reading, MA: Addison-Wesley. MR0344384
[2] Bedard, M. (2007). Weak convergence of Metropolis algorithms for non-iid target distributions. *Ann. Appl. Probab.* **17** 1222–1244. MR2344305
[3] Breyer, L.A. and Roberts, G.O. (2000). From Metropolis to diffusions: Gibbs states and optimal scaling. *Stochastic Process. Appl.* **90** 181–206. MR1794535
[4] Fang, K.T., Kotz, S. and Ng, K.W. (1990). *Symmetric Multivariate and Related Distributions. Monographs on Statistics and Applied Probability* **36**. London: Chapman and Hall. MR1071174
[5] Gelman, A., Roberts, G.O. and Gilks, W.R. (1996). Efficient Metropolis jumping rules. In *Bayesian Statistics, 5 (Alicante, 1994)* 599–607. New York: Oxford Univ. Press. MR1425429
[6] Krzanowski, W.J. (2000). *Principles of Multivariate Analysis: A User's Perspective*, 2nd ed. *Oxford Statistical Science Series* **22**. New York: The Clarendon Press Oxford Univ. Press. MR1133626
[7] Metropolis, N., Rosenbluth, A.W., Rosenbluth, M.N., Teller, A.H. and Teller, E. (1953). Equations of state calculations by fast computing machine. *J. Chem. Phys.* **21** 1087–1091.
[8] Roberts, G.O. (1998). Optimal metropolis algorithms for product measures on the vertices of a hypercube. *Stochastics Stochastic Rep.* **62** 275–283. MR1613256
[9] Roberts, G.O., Gelman, A. and Gilks, W.R. (1997). Weak convergence and optimal scaling of random walk Metropolis algorithms. *Ann. Appl. Probab.* **7** 110–120. MR1428751
[10] Roberts, G.O. and Rosenthal, J.S. (2001). Optimal scaling for various Metropolis–Hastings algorithms. *Statist. Sci.* **16** 351–367. MR1888450
[11] Sherlock, C. (2006). Methodology for inference on the Markov modulated Poisson process and theory for optimal scaling of the random walk Metropolis. Ph.D. thesis, Lancaster University. Available at http://eprints.lancs.ac.uk/850/.